\documentstyle[prd,aps,floats,graphicx]{revtex}

\begin{document}
\preprint{astro-ph/0207366}
\draft

%
%
\input epsf
\renewcommand{\topfraction}{0.8}
\twocolumn[\hsize\textwidth\columnwidth\hsize\csname
@twocolumnfalse\endcsname

\title{Effect of halo modeling on WIMP exclusion limits}
\author{Anne M.~Green} 
\address{Physics Department, Stockholm University, Stockholm, S106 91, Sweden} 
\date{\today} 
\maketitle
\begin{abstract}
WIMP direct detection experiments are just reaching the sensitivity
required to detect galactic dark matter in the form of
neutralinos. Data from these experiments are usually analyzed under
the simplifying assumption that the Milky Way halo is an isothermal
sphere with maxwellian velocity distribution. Observations and
numerical simulations indicate that galaxy halos are in fact triaxial
and anisotropic. Furthermore, in the cold dark matter paradigm
galactic halos form via the merger of smaller subhalos, and at least
some residual substructure survives.  We examine the effect of halo
modeling on WIMP exclusion limits, taking into account the detector
response. Triaxial and anisotropic halo models, with parameters
motivated by observations and numerical simulations, lead to
significant changes which are different for different experiments,
while if the local WIMP distribution is dominated by small scale
clumps then the exclusion limits are changed dramatically.

\end{abstract}


\vskip2pc]

\section{Introduction}

Arguably the best motivated non-baryonic dark matter candidate is the
neutralino (the lightest supersymmetric particle), and current direct
detection experiments are just reaching the sensitivity required to
probe the relevant region of parameter space~\cite{lars}.  The most
stringent exclusion limits on Weakly Interacting Massive Particles
(WIMPs) in general currently come from the Edelweiss~\cite{edelnew}
and Cryogenic Dark Matter Search (CDMS) experiments~\cite{CDMS}, with
competitive constraints also having been produced by
Heidelberg-Moscow~\cite{HM} and IGEX~\cite{IGEX}. The sensitivity to
WIMPs will be improved significantly in the short term future by the
continued operation of Edelweiss, and CDMS moving in a low background
environment at the Soudan mine~\cite{CDMS2}, and in the longer term by,
for instance, the planned GENIUS project~\cite{genius}.

The direct detection event rate, and its energy distribution, depend
crucially on the WIMP speed distribution. Data analyzes nearly always
assume a standard smooth halo model with isotropic maxwellian velocity
distribution. The change in the expected signal has been calculated
for various non-standard halo models of varying degrees of
sophistication~\cite{anal,newevans,uk,sikdm}. For models which are
effectively close to maxwellian, while there may be a significant
change in the annual modulation and angular dependence of the signal,
the change in the mean (averaged over time and recoil direction)
differential event rate is typically small~\cite{anal}. Models with
triaxiality or velocity anisotropy may however produce a significant
change even in the mean differential event rate~\cite{newevans,uk}.
Furthermore all of the non-standard halo models which have
previously been considered are essentially smooth\footnote{An
exception is Sikivie's late infall model~\cite{sik}, which contrary to
the standard picture of halo formation in CDM cosmologies (see
e.g. Refs.~\cite{mooredm,hws}) assumes axial symmetry and cold
collapse.}. N-body simulations, however, produce dark matter halos
which as well as being triaxial with anisotropic velocity
distributions~\cite{mooredm,hws} also contain
substructure~\cite{Nbody1}. A number of groups have recently
investigated the local dark matter distribution
numerically~\cite{swf,mooredm,hws}, using different methods and
reaching, to some extent, different conclusions.

Triaxiality, anisotropy and clumping in the WIMP velocity distribution
could potentially have a significant effect on the WIMP direct
detection signal.  Constraints (and in the future possibly best fits)
calculated assuming a standard maxwellian halo could be
erroneous~\cite{uk}. On the other hand, more optimistically, it might be
possible to derive useful information about the local velocity
distribution, and hence the formation of the galactic halo, if WIMPs
were detected~\cite{swf,hws}. Belli et. al.~\cite{damare} have
recently reanalyzed the DAMA collaborations annual modulation
signal~\cite{dama} for a range of halo models, finding that the
allowed region of WIMP mass--cross-section parameter space is
significantly enlarged. This illustrates that it is important to take
into account uncertainties in halo modeling when comparing exclusion
limits and/or allowed regions from different experiments.

Given the importance of the local dark matter distribution for direct
detection experiments we devote Sec.~\ref{halodis} to a detailed
discussion of the global properties of real and simulated dark matter
halos and recent work on the local dark matter
distribution~\cite{swf,mooredm,hws}.  In Sec.~\ref{analysis} we 
examine the effect of realistic halo modeling on exclusion limits. We
first investigate triaxial and anisotropic halos models, with
parameter choices motivated by the observations and simulations, and
then, more speculatively examine the possible effects of small
subhalos.

\section{Galaxy halos}
\label{halodis}

\subsection{Global properties}

Observational constraints on the structure of dark matter halos depend
on the relation of luminous tracer populations to the underlying
density distribution, and are complicated by galactic structure and
projection effects. Ref.~\cite{sackett} concludes that in the
outskirts of spiral galaxies the intermediate-to-long axis ratio is
likely to be greater than $0.8$, while the short-to-long axis ratio is
largely unconstrained with values in the range $0.3-1.0$ reported,
with some correlation between the method used and the value obtained
(see Refs.~\cite{sackett} and \cite{om2} for details and
references). For the Milky Way (MW), analysis of local stellar
kinematics gives an estimate for the short-to-long axis value of $0.7
\pm 0.1$~\cite{om2}, while the great circle tidal streams observed
from the Sagittarius dwarf galaxy rule out ratios of less than 0.7 in
the outer halo at high confidence~\cite{dsg} (in a flattened potential
angular momentum is not conserved, so that orbits precess and tidal
streams lose their coherence). The (an)isotropy of velocities in the
MW halo is even harder to probe, however there is some evidence that
galactic globular clusters may have preferentially radial
orbits~\cite{glob}.

Given the difficulties involved in `observing' galaxy halos it makes
sense to turn to numerical simulations for information on their
possible structure. Current simulations of galaxy halos within a
cosmological context can resolve sub-kpc scales (see
e.g.~\cite{Nbody1,Nbody2}). Discrepancies between the halos produced
in these simulations (which have lots of surviving dwarf galaxy sized
subhalos and steep central profiles) and observations, have led to
claims of a crisis for the cold dark matter model (see
Ref.~\cite{cdmcrisis} and references therein for extensive
discussion). Most relevant for the local dark matter distribution is
the subhalo problem which may be, at least partly, due to
complications in comparing the observed luminous matter with the dark
matter distribution from the simulations. In particular it has been
argued that gas accretion onto low mass halos may be inhibited after
reionization so that a large fraction of the subhalos remain
dark~\cite{sf}. It has also been shown that if the observed dwarf
galaxies themselves have dark halos, then their masses have been
underestimated and correcting for this would go toward resolving the
discrepancy with observations~\cite{hay,dg}. The survival of subhalos
is at least partly due to their concentrated profiles, so any
modification to the simulations which produced halos with shallower
central profiles could also reduce the number of surviving
subhalos. Despite the ongoing debate regarding the detailed comparison
of the small scale properties of simulated halos with observations,
cosmological simulations may still provide us with useful information
about the global properties of dark halos.

The shape of simulated halos varies, not just between different halos
of the same mass, but also as a function of radius within a single
halo, strongly if the halo has undergone a major merger relatively
recently. Two high resolution Local Group halos studied in detail in
Ref.~\cite{mooredm} have axis ratios of $1:0.78:0.48$ and
$1:0.45.0.38$ at the solar radius and $1:0.64:0.40$ and $1:0.87:0.67$
at the virial radius. Adding dissipative gas to simulations tends to
preserve the short-to-long axis ratio while increasing the
intermediate-to-long axis ratio~\cite{gas}.

The anisotropy parameter $\beta(r)$, defined as
\begin{equation}
\label{defbeta}
\beta(r)= 1 - \frac{<v_{ \theta}^2>+<v_{ \phi}^2> }{2 <v_{{\rm r}}^2>} \,,
\end{equation}
where $<v_{{\rm \theta}}^2>$, $<v_{{\rm \phi}}^2>$ and $<v_{{\rm
r}}^2>$ are the means of the squares of the velocity components
evaluated at radius $r$, also varies with radius.  Typically
$\beta(r)$ grows, although not monotonically, from roughly zero in the
center of the halo to close to one at the virial radius, with
non-negligible variation between halos (see Fig. 23 of
Ref.~\cite{fm}). The high resolution galactic mass halos studied in
Ref.~\cite{moore99} have $\beta(R_{\odot})$ in the the range 0.1-0.4,
corresponding to radially biased orbits.

\subsection{Local dark matter distribution}

In CDM cosmologies structure forms hierarchically, from the top
down~\cite{cdmgen}. Small objects (often known as subhalos) form
first, with larger objects being formed progressively via mergers and
accretion. The internal structure of large galaxy size halos is
determined by the dynamical processes which act on the accreted
components. Dynamical friction causes subhalos with mass $M \gtrsim
10^{9} M_{\odot}$ to spiral toward the center of their parent halo
within a Hubble time~\cite{bt}, while the tidal field of the main halo
can strip material away from a subhalo~\cite{bt,hay} producing tidal
streams along its orbit~\cite{ts}.

The local dark matter distribution can not be probed directly by
cosmological simulations; the smallest subhalos resolvable in the
highest resolution simulations have mass of order $10^{7} M_{\odot}$
and it is not possible to fully resolve substructure within
subhalos. Little substructure is found within the central regions of
simulated halos, however it is not known whether the subhalos have
been destroyed by tidal stripping or if this is purely a resolution
effect~\cite{mooredm}. This is crucial for the local dark matter
distribution as the solar radius ($R_{\odot} \approx 8$ kpc) is small
compared with the radius of the MW halo, which from observations is
thought to be in excess of 100 kpc~\cite{sackett}, while simulated
halos with the same peak circular velocity as the MW have virial
radii\footnote{The virial radius is the radius which separates
the virialized and infall regions of simulated halos and has a mean
density within it of 178 (100) times the critical density at $z=0$ in
the standard($\Lambda$)CDM cosmology~\cite{rvir}.}  of order 200 kpc.
The computing power required to directly probe the local dark matter
distribution will probably not be available for a decade or
so~\cite{mooredm}, therefore other numerical~\cite{mooredm,hws} and
semi-analytic~\cite{swf} approaches have been used to address the
problem.

\vspace{0.5cm}

Stiff, Widrow and Frieman~\cite{swf} employ a semi-analytical
approach, calculating the subhalo distribution as a function of mass
and accretion redshift using the extended Press-Schechter
formalism~\cite{eps} and then evolving individual late accreting
smooth subhalos within a growing spherical halo, to find the
probability distribution of the over-density at the solar radius. They
find that there is a high (of order one) probability that there is a
density enhancement of $\sim 3 \%$ of the mean halo density in the
solar neighborhood, and the probability of an enhancement roughly
equal to the background density is non-negligible (of order
0.01)\footnote{These probabilities are lower limits as they only
include the contributions of subhalos accreted after $z=1$.}.  They
show that if the velocity of the clump, or stream, with respect to the
Earth is high enough it will produce a shoulder in the differential
event rate at high energies.

\vspace{0.5cm}

Moore et. al.~\cite{mooredm} take a region with dynamical properties
similar to the Local Group, resimulated at higher resolution from a
standard CDM cosmological simulation, and identify a subhalo with mass
similar to the Draco dwarf galaxy. They then resimulate this subhalo
up until its merger with the parent halo, at which point roughly
10$\%$ of its mass is in the form of (sub)subhalos. To assess the
effect of tidal friction, subhalos with smooth density profiles are
evolved within a smooth Galactic potential and it is found that small
subhalos orbiting at the solar radius which are accreted onto the
Galactic halo early, retain most of their mass due to their high
central densities.

Moore et. al. conclude that the phase space distribution at the solar
radius will depend crucially on the Galaxy's merger history and on the
internal structure of the smallest subhalos, arguing that it is
possible that the local dark matter density could be zero or that a
single dark matter stream with small velocity dispersion could
dominate or that many tidal streams could overlap to give a smooth
distribution. The solution to this problem depends on the extent to
which the substructure within subhalos is destroyed prior to their
accretion onto the main halo. The subhalos are of course also formed
hierarchically themselves from smaller subhalos. The free-streaming
length of neutalinos is so small that the first clumps to form have
mass $M \sim 10^{-12} M_{\odot}$\cite{hss}, however, and to follow the
accretion and destruction of subhalos though such a large hierarchy of
scales would be decidedly non-trivial.

\vspace{0.5cm}

Helmi, White and Springel~\cite{hws} take a cosmological $\Lambda$CDM
simulation, where the second largest cluster has been resimulated at
higher resolution and then scaled down in size to match the MW. Most
of the mass in the inner halo has been in place for 10 Gyr and the
smallest resolvable subhalos do not survive to the present day at the
solar radius. They argue that the only way a small subhalo with high
density could exist in the solar neighborhood at $z=0$ is if it were
first accreted by a large halo which subsequently had a major merger
with the main progenitor of the Galactic halo. They conclude that the
local dark matter velocity distribution is well approximated by a smooth
multi-variate gaussian, with clumps of high velocity particles present
if the MW halo has undergone a recent major merger.

\vspace{1.0cm}

In summary numerical simulations produce galaxy halos which are
significantly triaxial and anisotropic, with the shape and anisotropy
of a halo depending on its individual merger history, a picture which
is broadly supported by observations. This indicates that, even if the
local velocity distribution is relatively smooth, the standard
spherical isotropic maxwellian halo model may not be a good
approximation.  Furthermore galaxy halos are formed hierarchically
from the accretion of smaller subhalos, which may not be completely
destroyed by tidal friction, so that the local dark matter could be
distinctly non-smooth. It is even possible that the dark matter could
be distributed largely in small dense clumps. There is currently no
consensus on the local dark matter velocity distribution however, with
the results obtained depending on the method used to extrapolate to
small scales below the resolution limit of cosmological simulations,
so this possibility should be regarded as speculative.

\section{Effects on experimental analysis}
\label{analysis}

The differential WIMP event rate due to scalar interactions can be
written in terms of the WIMP scattering cross section on the proton,
$\sigma_{{\rm p}}$~\cite{jkg}:
\begin{equation}
\frac{{\rm d} R}{{\rm d}E} = \zeta \sigma_{{\rm p}} 
              \left[ \frac{\rho_{0.3}}{\sqrt{\pi} v_{0}}
             \frac{ (m_{{\rm p}}+ m_{\chi})^2}{m_{{\rm p}}^2 m_{\chi}^3}
             A^2 T(E) F^2(q) \right] \,,
\end{equation}
where the local WIMP density, $\rho_{\chi}$ is normalized to a
fiducial value $\rho_{0.3} =0.3 \, {\rm GeV/ cm^{3}}$, such that
$\zeta=\rho_{\chi} / \rho_{0.3}$, $m_{A}$ is the atomic mass of the
target nuclei, $E$ is the recoil energy of the detector nucleus, and
$T(E)$ is defined as~\cite{jkg}
\begin{equation}
\label{tq}
T(E)=\frac{\sqrt{\pi} v_{0}}{2} \int^{\infty}_{v_{{\rm min}}} 
            \frac{f_{v}}{v} {\rm d}v \,,
\end{equation}
where $f_{v}$ is the WIMP speed distribution in the rest frame of the
detector, normalized to unity, and $v_{{\rm min}}$ is the minimum
detectable WIMP speed
\begin{equation}
v_{{\rm min}}=\left( \frac{ E (m_{\chi}+m_{A})^2}{2 m_{\chi}^2 m_{A}} 
             \right)^{1/2} \,.
\end{equation}

\subsection{Triaxiality and anisotropy}

We will first examine the change in the WIMP speed distribution for
triaxial and anisotropic, but still smooth, halo models. To date two
self-consistent triaxial and/or anisotropic halo models have been
studied in relation to WIMP direct detection: the logarithmic
ellipsoidal model~\cite{newevans} and the Osipkov-Merritt anisotropy
model~\cite{OM}, studied in Ref.~\cite{uk}. We will extend the
previous work by focusing on parameters which span the range of halo
properties discussed in Sec.~\ref{halodis} above and including the
detector response.

\begin{table}
\label{tab1}
\begin{center}
\begin{tabular}{|c|c|c|}
$\beta$ & $p=0.9, q=0.8$ & $p=0.72, q=0.7$ \\ 
\hline
\multicolumn{3}{c} {\rm
intermediate axis} \\ 
\hline
0.1 & 0.07 & 4.02 \\ 0.4 & -0.62 & 2.01 \\
\hline
\multicolumn{3}{c} {\rm major axis} \\ 
\hline
0.1 & -1.00 & -1.39 \\ 0.4 &
-1.33 & -1.60 \\
\end{tabular}
\end{center}
\caption[g]{\label{g} The values of the anisotropy parameter $\gamma$
in the logarithmic ellipsoidal model required to produce $\beta=0.1$
and 0.4.}
\end{table}

\begin{figure}[t]
\centering
\includegraphics[width=0.45\textwidth]{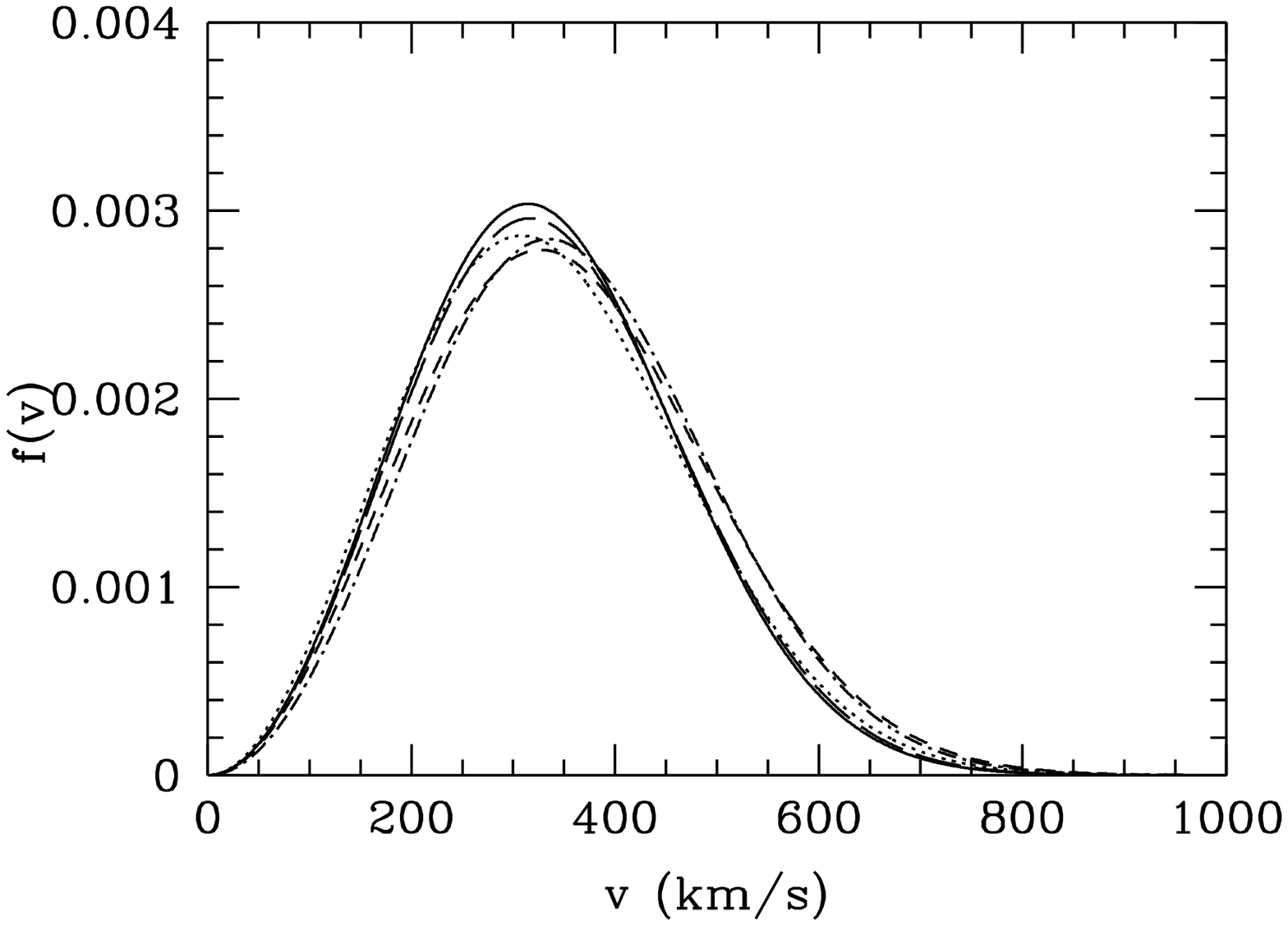}
\includegraphics[width=0.45\textwidth]{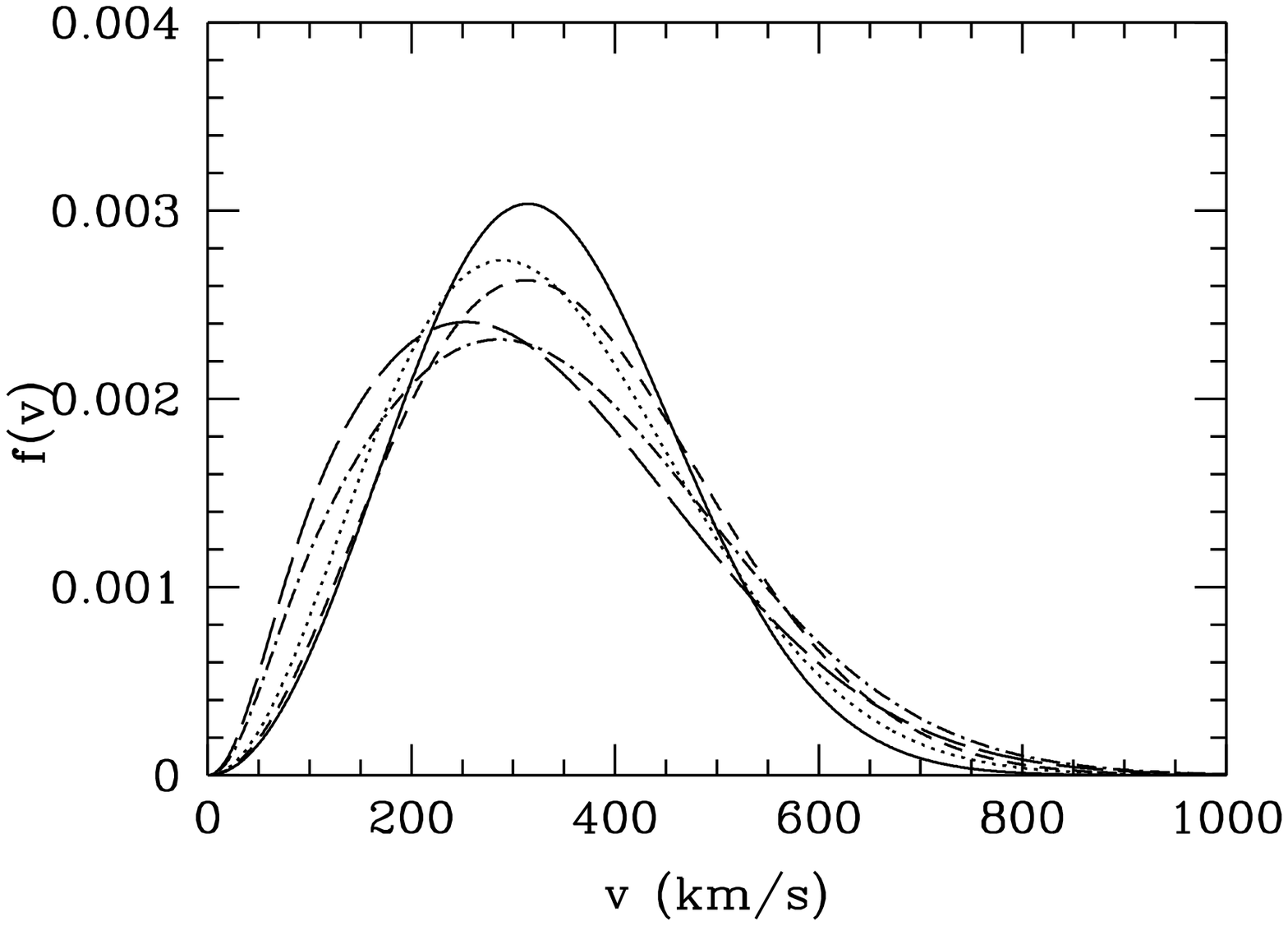}
\caption[fig1]{\label{fig1} The speed distributions, in the rest frame
of the Sun, for the standard halo model (solid line), and triaxial
models on the major axis (upper panel) and intermediate axis (lower
panel) for p=0.9, q=0.8 and $\beta=0.1/0.4$ (dotted/short dashed) and
for p=0.72, q=0.7 and $\beta=0.1/0.4$ (long dashed/dot dashed).}
\end{figure}

The logarithmic ellipsoidal model~\cite{newevans} is the simplest
triaxial generalization of the isothermal sphere and the velocity
distribution can be approximated by a multi-variate gaussian on either
the long or the intermediate axis (see Appendix A for
further details)\footnote{Of course there is no reason to expect the Sun
to be located on one of the axes of the halo.}.  We consider parameter
values $p= 0.9, q=0.8$ corresponding to axial ratios $1:0.78:0.48$ (as
used in Ref.~\cite{newevans}), and $p=0.72, q=0.70$ corresponding to
$1:0.45:0.38$, and locations on the long and intermediate axes. The
first set of axial ratios is typical of the values found in
simulations and roughly consistent with observations, while the second
is arguably rather extreme. We use values of the anisotropy parameter
$\gamma$ (which in the spherical limit $p=q=1$ is related to $\beta$
by $-\gamma= 2 \beta$) which give $\beta=0.1$ and $0.4$, and are
tabulated in Table I. The speed distributions, in the rest frame of
the Sun normalized to unity, are plotted in Fig.~\ref{fig1} along
with that for the standard maxwellian halo model. For both positions
the triaxial models have a wider spread in speeds than the standard
model, so that the differential event rate will decrease less rapidly
with increasing recoil energy, but the change is small on
the major axis. This is because the change in the speed distribution
is largely determined by the velocity dispersion in the $\phi$
direction.  On the major axis, for parameter values which give $0.1 <
\beta < 0.4 $, all three components of the velocity have roughly the
same dispersion, whereas on the intermediate axis the velocity
dispersion in the $\phi$ direction is significantly larger than that
in the $z$ direction. Note that the speed distributions we consider
deviate less from the standard maxwellian than those considered by
Refs.~\cite{newevans} and~\cite{damare} as they use more extreme
values for $\gamma$ (namely 16 and -1.78).

\begin{figure}
\centering
\includegraphics[width=0.45\textwidth]{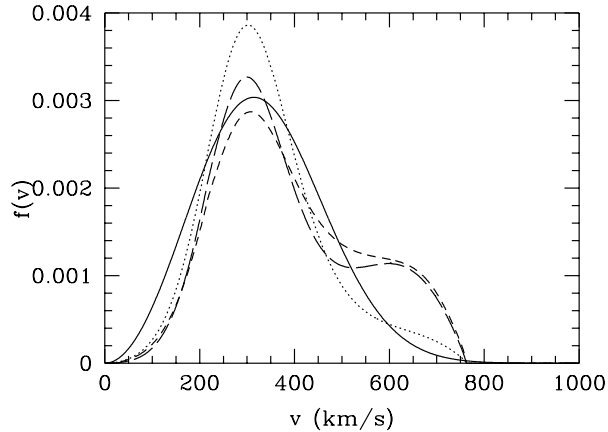}
\caption[fwid]{\label{fwid} The speed distributions for the standard
halo model (solid line), and the OM anisotropy model with $\beta=0.13,
 0.31$ and $0.4$ (dotted, short-dashed, and long-dashed).}
\end{figure}

In the Osipkov-Merritt (OM) model (see Appendix B for further
details), which assumes a spherically symmetric density profile, the
velocity anisotropy varies as a function of radius as
\begin{equation}
\beta(r) = \frac{r^2}{r^2 + r_{{\rm a}}^{2}} \,,
\end{equation}
so that the degree of anisotropy increases with increasing radius, as
is found in numerical simulations. Following Ref.~\cite{uk} we assume
a NFW~\cite{NFW} density profile\footnote{Varying the inner slope of
the density profile does not significantly affect the local velocity
distribution~\cite{uk}.} with scale radius $r_{{\rm s}}=20$ kpc. We
use values of the anisotropy radius $r_{{\rm a}}=20, 12,$ and $9.8$
kpc which correspond to $\beta(R_{0}) = 0.14, 0.31$ and $0.4$
respectively. For the first two values analytic fitting functions for
the distribution function have been provided by
Widrow~\cite{widrow}. The resulting speed distributions are plotted in
Fig.~\ref{fwid} along with that for the standard maxwellian halo
model. The excess at large $v$ is due to the increased number of
particles on very elongated, nearly radial orbits~\cite{uk}.

\begin{figure}[t]
\centering
\includegraphics[width=0.45\textwidth]{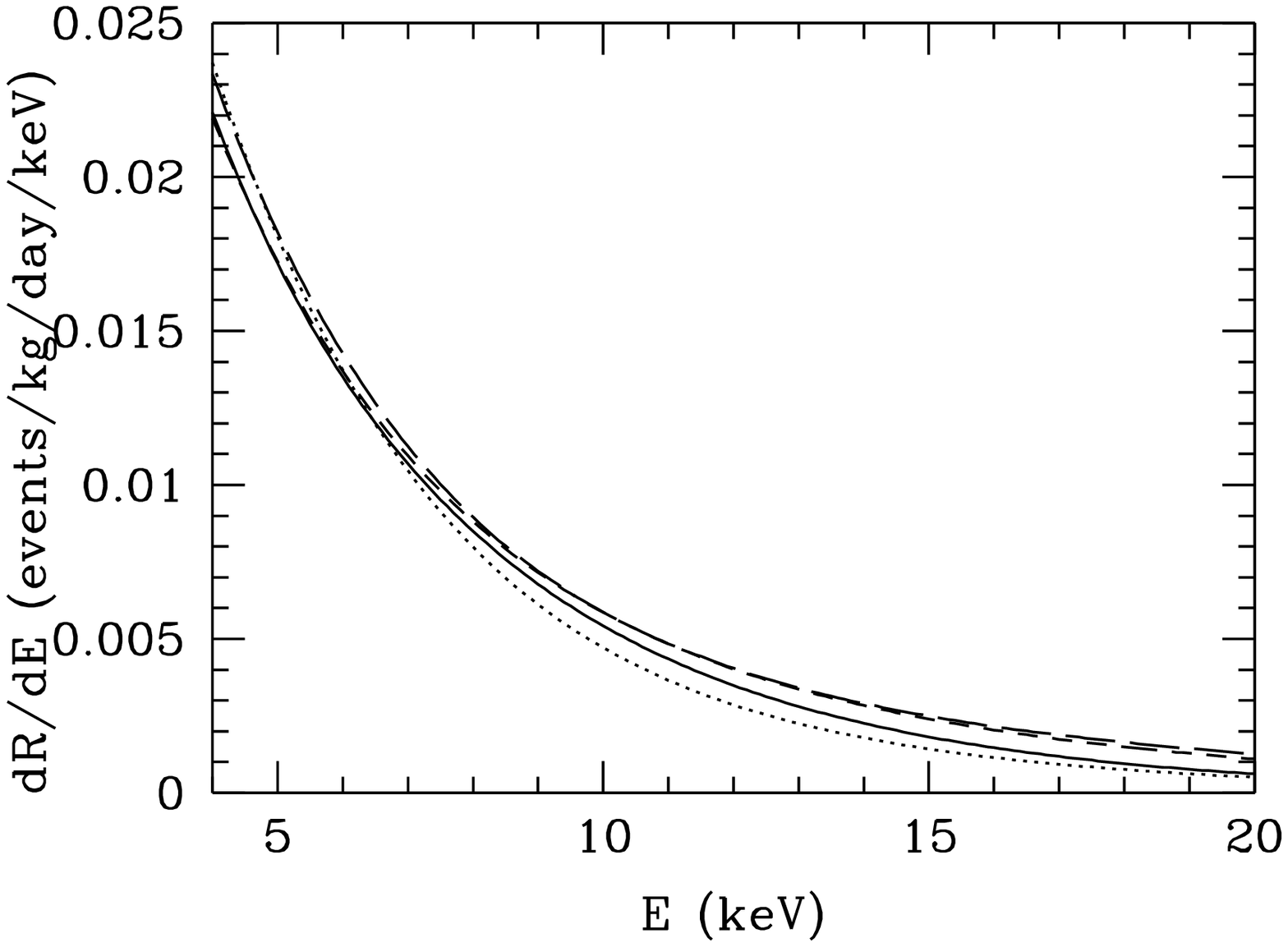}
\includegraphics[width=0.45\textwidth]{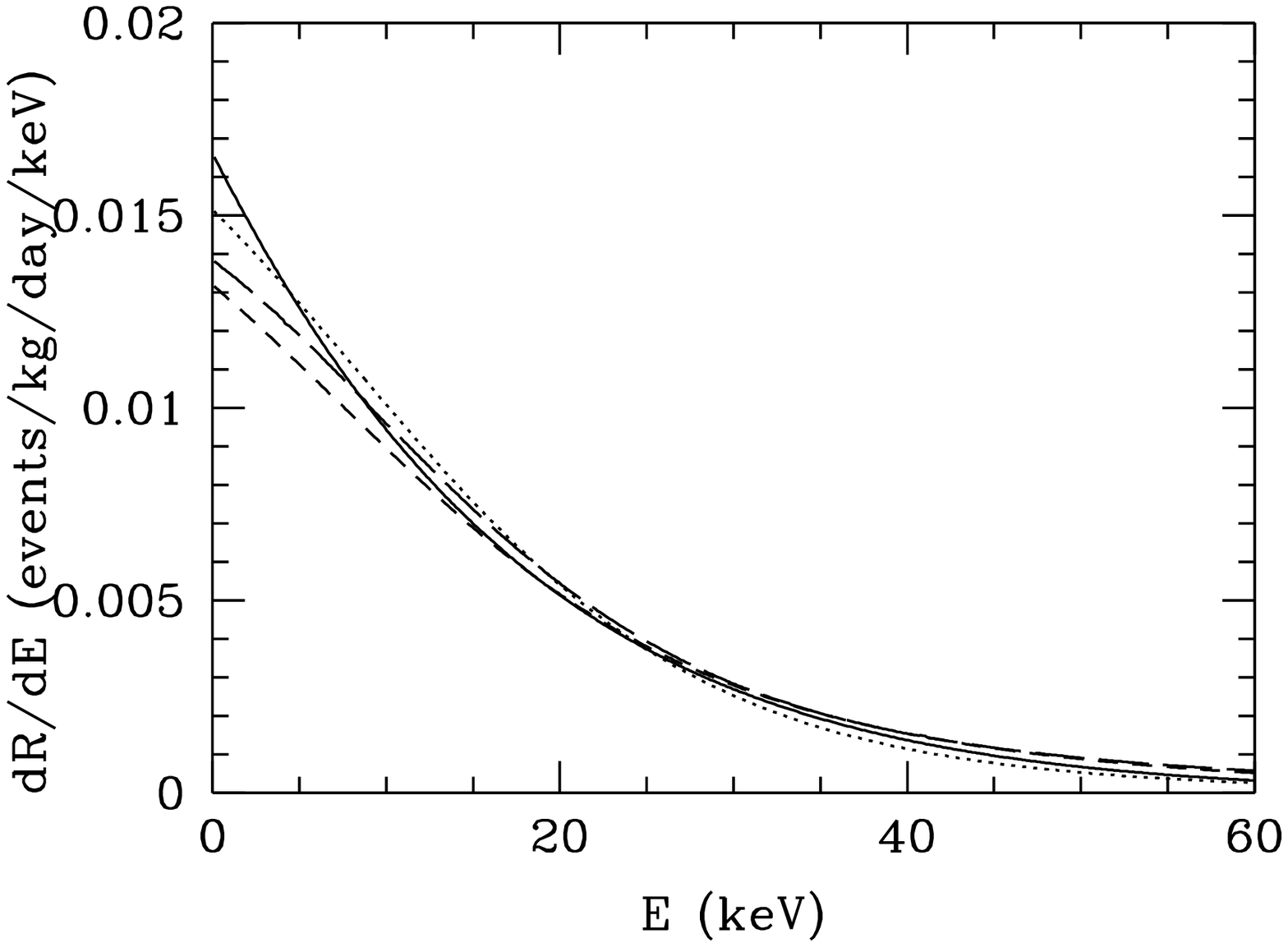}
\caption[drde]{\label{drde} The differential event rate for the OM
anisotropy model, with speed distributions as plotted in Fig.~2, for
$m_{{\chi}}=50$ GeV and $\sigma_{{\rm p}}= 10^{-45} {\rm m^{2}}$for
the IGEX detector (upper panel) and for an `ideal' Ge detector
(lower panel).}
\end{figure}

 To assess the effect of changes in the speed distribution on
exclusion limits we need to take into account the detector response,
including the difference between the observed energy of an event and
the actual recoil energy, non-zero energy threshold and energy
resolution (see Ref.~\cite{ls} for further details), as these factors
may blur out the effects of changes in the speed distribution. We
consider a ${\rm Ge}^{76}$ detector with energy threshold $E_{{\rm
T}}=4$ keV with the same properties (resolution and form factor) as
that used by the IGEX experiment~\cite{IGEX}, which is optimized for
detecting double-beta decay.  The resulting differential event rates,
per kg per day per keV, are plotted in Fig.~\ref{drde}, for the OM
speed distributions in Fig.~\ref{fwid}, for WIMPs with mass
$m_{\chi}=50$ GeV and cross-section $\sigma_{{\rm p}}= 10^{-45} {\rm
m^{2}}$. We see that, in this model, the differential event rate does
not deviate linearly from that of the standard halo model as the
degree of anisotropy is increased.  Since future detectors, optimized
for WIMP detection, will have lower thresholds and better energy
resolution, we also plot the differential event rates for an `ideal'
(i.e. completely unrealistic) $A=76$ detector where the full recoil
energy is detected, the energy resolution is perfect and $E_{{\rm
T}}=0$ keV~\footnote{Somewhat counter-intuitively, at low energies the
differential event rate for the IGEX detector is higher than that for
the ideal detector, due to the finite energy resolution of the IGEX
detector.}. The difference between the differential event rates is
then largest at small recoil energies, and would therefore be more
significant for an experiment with a lower threshold energy.

\begin{figure}[t]
\centering
\includegraphics[width=0.45\textwidth]{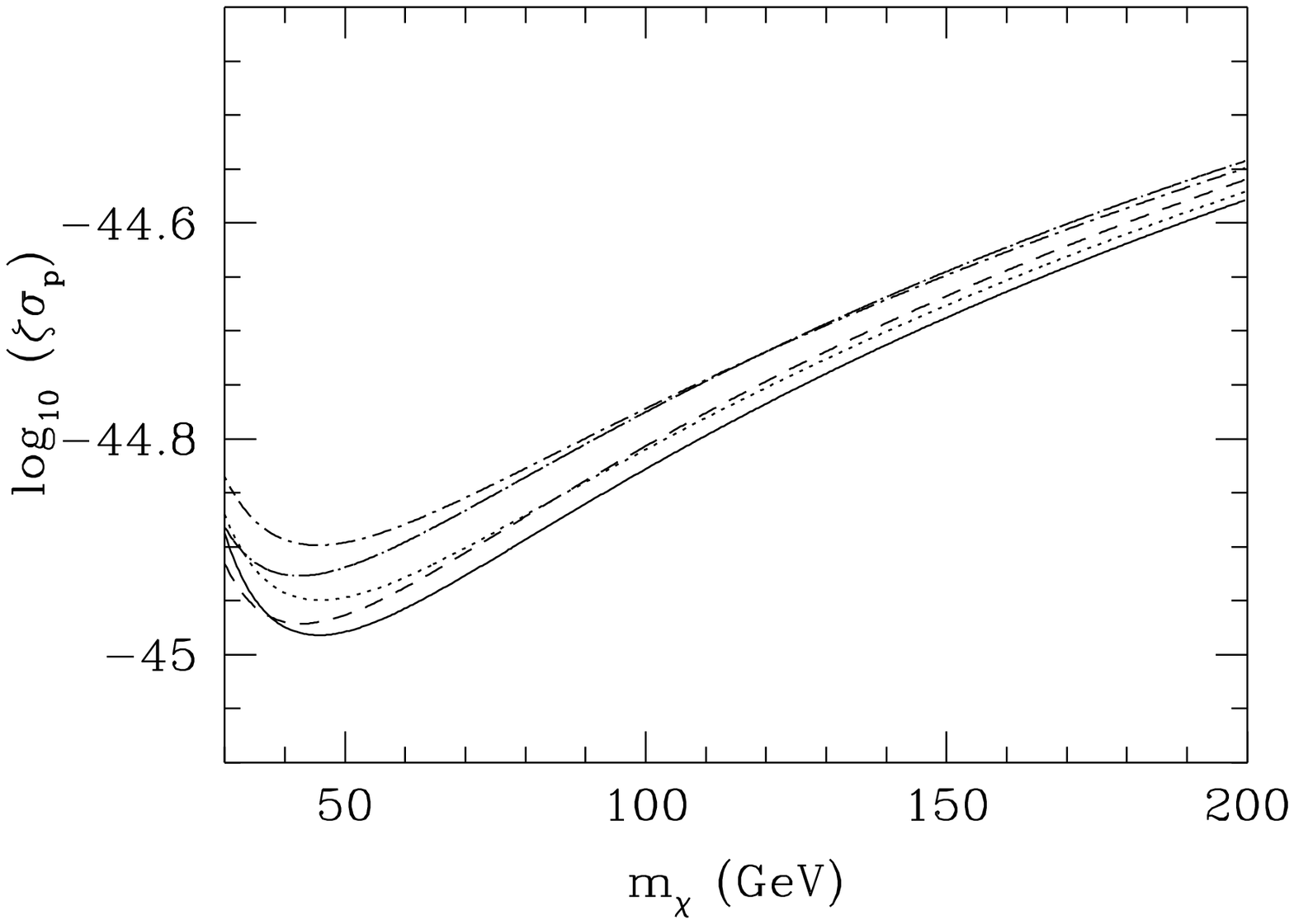}
\includegraphics[width=0.45\textwidth]{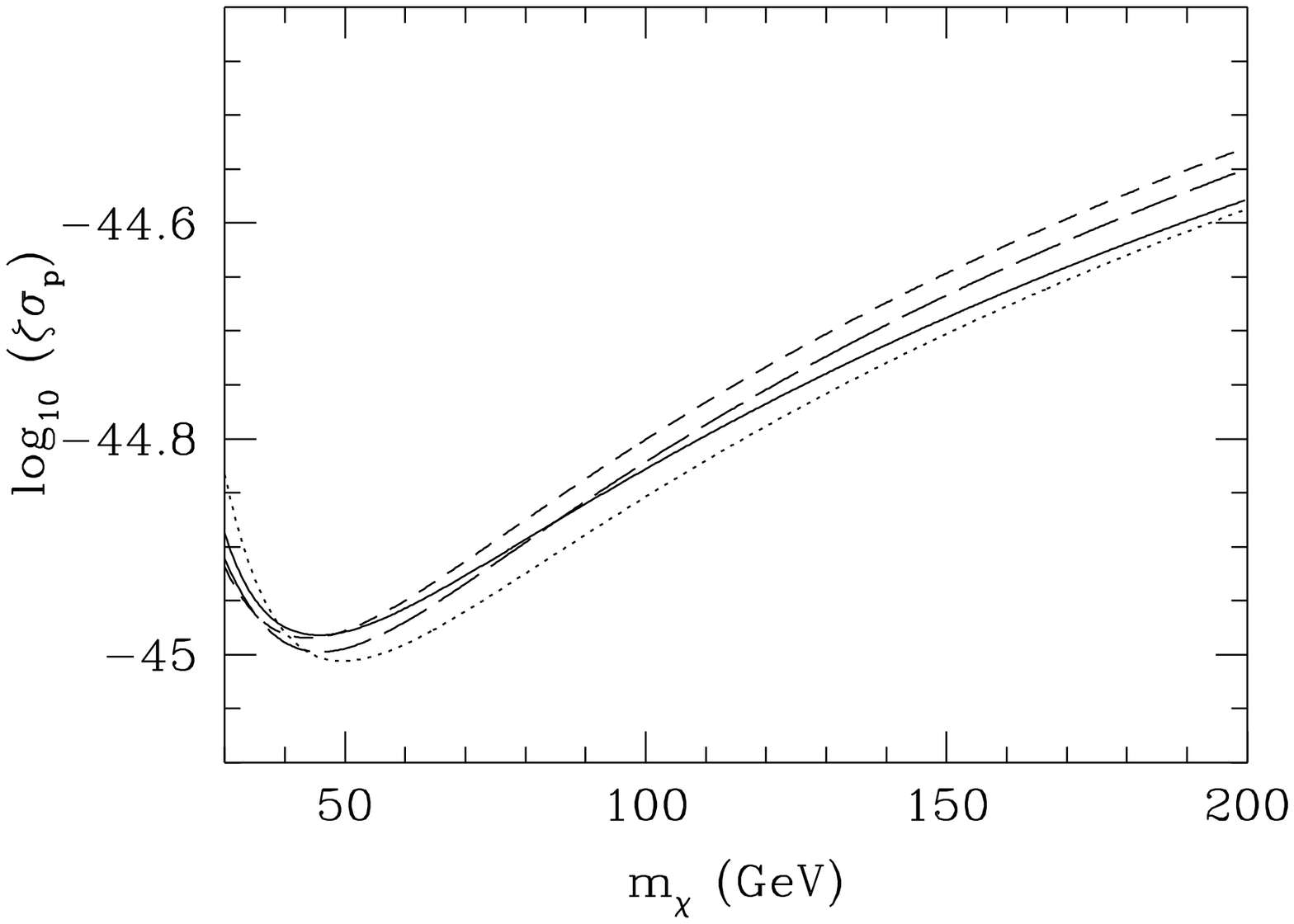}
\caption[IGEX]{\label{IGEX} The exclusion limits from the IGEX
experiment for the logarithmic ellipsoidal model, location on the
intermediate axis (upper panel) and for the OM model
(lower panel). Line types as in the lower panel of Fig. 1 and Fig. 2
respectively.}
\end{figure}

\begin{figure}
\centering
\includegraphics[width=0.45\textwidth]{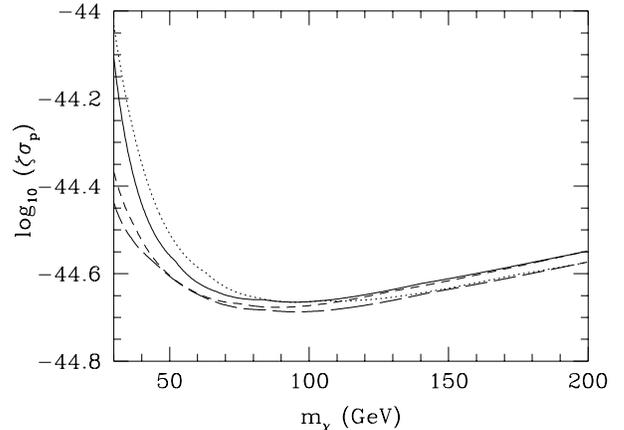}
\caption[HM]{\label{HM} The exclusion limits from the HM experiment
for the OM model. Lines as in Fig. 2.}
\end{figure}

In Fig.~\ref{IGEX} we plot the exclusion limits found from the IGEX
data by requiring that the data in no more than one energy bin exceeds
its 99.77$\%$ confidence limit, so as to produce 90$\%$ overall
confidence limits~\cite{statpap}, for the logarithmic ellipsoidal
model and for the OM anisotropy model. We also plot the exclusion
limits from the Heidelberg-Moscow (HM) experiment~\cite{HM} for the OM
anisotropy model in Fig.~\ref{HM}. Comparing Figs.~\ref{IGEX} and
~\ref{HM} we see that the change in the exclusion limits depends not
only on the halo model under consideration, but also on the data being
used; for IGEX the change in the exclusion limits is largest for large
WIMP masses, while for HM the change is largest for small WIMP
masses. For different WIMP masses, different energy ranges can be most
constraining; for the IGEX data the lowest energy bin is always the
most constraining, while for HM as the WIMP mass increases the
constraint comes from higher energy bins. It should therefore be borne
in mind when comparing exclusion limits from different experiments,
that changing the assumed WIMP speed distribution will affect the
limits from different experiments differently.

The change in the exclusion limits is not huge (of order tens of
per-cent) for the experiments we have considered, however these
experiments are not optimized for WIMP detection. As illustrated in
Fig.~\ref{drde} the change in the differential event rate, and hence
the exclusion limit, would be significantly larger for an experiment
with better energy resolution and lower threshold energy. We have also
seen that different models with the same value for the anisotropy
parameter $\beta$ have very different speed distributions, and hence a
different effect on the exclusion limits. Furthermore it is conceivable
that the local WIMP velocity distribution may deviate even further
from the standard maxwellian distribution, than the models that we
have considered.

\subsection{Clumps}

 Even if dynamical processes produce a smooth background dark matter
distribution, late accreting clumps may lead to a local density
enhancement and velocity clumping~\cite{swf,hws}, and produce a
shoulder in the differential event rate, if their density and velocity
with respect to the earth are large enough.  For the experiments we
have been considering the lower energy bins are most constraining, so
that only very rare high density and velocity streams would have a
non-negligible effect on the exclusion limits. The effect of these
late accreting clumps on the annual modulation and directional signals
would be more significant however~\cite{swf}.

\begin{figure}
\centering
\includegraphics[width=0.45\textwidth]{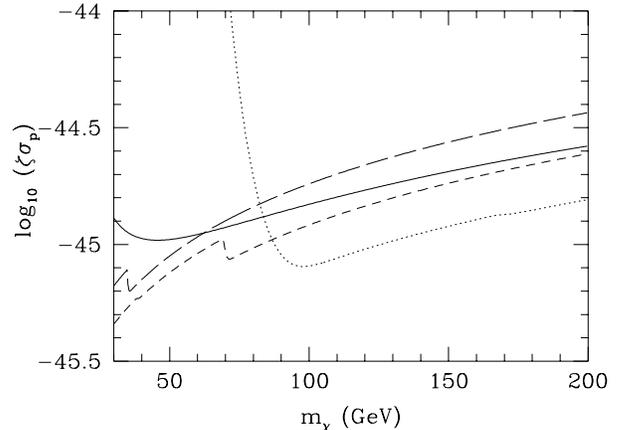}
\caption[step]{\label{step} The exclusion limits from the IGEX data if
a subhalo with negligible velocity distribution moving with speed 200,
400, 600 ${\rm km s^{-1}}$ relative to the earth (dotted, short
dashed, long dashed) dominates the local density.}
\end{figure}

We will now turn our attention to the consequences of the more
speculative possibility that small subhalos may survive at the solar
radius. We could then be located within a subhalo with local density
in excess of the mean value of $0.3 \, {\rm GeV cm^{-3}}$, on the
other hand it is even possible that we could be in a region between
clumps and streams where the WIMP density is zero~\cite{mooredm}. In
the latter case all attempts at WIMP direct detection would be doomed
to failure, and exclusion limits would tell us nothing about the WIMP
cross-section. At the other extreme a tiny subhalo at the earth's
location would produce a distinctive signal and, due to the enhanced
density, make it easier to detect WIMPs of a given
cross-section. Subhalos with $M \ll 10^{9} M_{\odot}$ would have
negligible velocity dispersion and hence a delta-function speed
distribution. The resulting theoretical differential event rate would
be a step function with amplitude inversely proportional to the speed
of the subhalo with respect to the earth, and position increasing with
increasing relative speed and WIMP mass. Consequently for small subhalo
velocities and WIMP masses there would be no constraint on the WIMP
cross-section (no WIMPs would have sufficient energy to cause an
observable recoil), but as the WIMP mass is increased the constraints
would become much tighter as then all the WIMPs would be energetic
enough to cause events of a given recoil energy.  In Fig.~\ref{step}
we plot the exclusion limits on $\zeta \sigma_{\rm p}$ from the IGEX
data for subhalos with various relative speeds\footnote{In reality the
subhalo speed would be another unknown variable.} dominating the local
WIMP distribution. Unlike the smooth halo models for this data set, as
the WIMP mass is increased higher energy bins are most constraining,
and this leads to sharp changes in the exclusion limits due to the
sharp transition from no WIMPs to all WIMPs having sufficient to
energy cause events of a given recoil energy. Note that the high
density of a subhalo would lead to $\xi > 1$, so that the exclusion
limits on the WIMP cross-section would be tighter (by some unknown
factor) than for the standard halo model.

How dense could small scale clumps be? Dark matter clump densities are
usually parameterized in terms of their concentration, $c$, defined as
\begin{equation}
c= \frac{R_{{\rm vir}}}{r_{{\rm s}}} \,,
\end{equation}
where $R_{{\rm vir}}$ is the virial radius and $r_{{\rm s}}$ is the scale
radius, the radius at which the effective logarithmic slope of the
density profile is equal to -2 (see e.g. Refs.~\cite{bull,ens}). Low
mass halos typically form earlier, when the density of the Universe is
higher, so that their concentrations are higher than those of larger
halos~\cite{NFW} and Bullock et. al.~\cite{bull} and Eke, Navarro and
Steinmetz (ENS)~\cite{ens} have recently constructed toy models which
reproduce the scaling of concentration with mass and redshift for the
simulated galaxy-sized halos.

Since the first neutralino clumps to form have mass of order $10^{-12}
M_{\odot}$~\cite{hss}, the first generation of halos which are formed
in the same way as galaxy sized halos (via the accretion and merger of
smaller clumps) will have mass of order $10^{-10} M_{\odot}$, and
virial radii of order $0.01$ pc. While these clumps are tiny by
cosmological standards, they are still relatively large compared to
the distance traveled by the Sun in a year ($10^{-4}$
pc). Extrapolating the Bullock et. al. and ENS toy models, way beyond
their intended range of applicability, produces significantly
different `guesstimates' for $c$ for $M \sim 10^{-10} M_{\odot}$ halos
of 150 and 30 respectively~\cite{ubel}. Subhalos within larger halos
may in fact be more concentrated than isolated halos of the same mass
as dense regions tend to collapse earlier and tidal stripping may
steepen their density profiles~\cite{bull}.  The tidal
radius~\cite{bt} of a $10^{-10} M_{\odot}$ subhalo orbiting in the MW
halo at the solar radius is of order $10^{-3}$ pc, so the dense
central regions of these subhalos should survive tidal
stripping~\cite{mooredm,hay}. The density of simulated halos diverges
toward their center, so as a measure of the typical density of a
subhalo we use the mean density within the scale radius, assuming a
NFW density profile~\cite{NFW}:
\begin{equation}
\rho(r) \propto \frac{r_{{\rm s}}}{r ( 1 + r/r_{{\rm s}})^2} \,.
\end{equation}
For $c=30 (150)$ this gives $\rho=0.9 (70) {\rm GeV cm^{-3}}$, or
equivalently $\zeta= 3 (200)$. We emphasize that this rough
calculation relies on the extrapolation of the concentration-mass
scaling models far beyond the regime in which they have been
numerically tested (isolated galaxy mass halos) and the numbers
obtained should probably not be taken seriously. It does illustrate,
however, that {\em if} small subhalos survive then they {\em may} lead
to a significant local enhancement in the WIMP density, and that
understanding the small scale structure of galactic halos is crucial
for realistic modeling of the WIMP velocity distribution.

\section{Discussion}
Rapid progress is being made in the field of WIMP direct detection,
with experiments closing in on the sensitivity required to detect
neutralinos, if they constitute a non-negligible fraction (greater
than $10^{-4}$) of the halo density~\cite{dgg}. Data analyzes
usually assume the simplest halo model: an isothermal sphere with
maxwellian velocity distribution. There is no clear justification,
either observational or theoretical, for this assumption apart from
simplicity. In fact numerical simulations~\cite{Nbody1,mooredm,hws}
and observations~\cite{sackett,om2,glob} suggest that galaxy halos are
triaxial and anisotropic. The local density distribution may also be
non-smooth, with late accreting subhalos leading to velocity
clumping~\cite{swf,hws}. More speculatively it is even possible that
the dark matter could be mainly distributed in tiny dense clumps, so
that the local density distribution could be dominated by a single
clump, or could even be zero~\cite{mooredm}. It is therefore crucial
to examine the effect of realistic halo modeling on the WIMP direct
detection signal.

In this paper we have investigated the change in exclusion limits due
to triaxiality, velocity anisotropy and small scale clumping, taking
into account detector performance and using parameter values motivated
by numerical simulations and observations. Triaxiality and velocity
anisotropy lead to non-negligible changes in the exclusion limits,
even for detectors with relatively poor energy resolution. Furthermore
the changes are different for different data sets and depend on how
the anisotropy is modeled. If the local WIMP distribution is
dominated by small scale clumps then the local density may be zero
(making it impossible to detect WIMPs) or significantly enhanced
(making it easier to detect WIMPs with a given cross-section), and the
exclusion limits are changed dramatically. Clearly the survival of
subhalos at the solar radius is a very important issue for WIMP direct
detection.

Even if the local WIMP distribution is smooth, to derive reliable
constraints on WIMP parameters and compare results for different
experiments a framework needs to be developed for dealing with the
uncertainty in the WIMP velocity distribution; either the development
of a framework for parameterizing deviations from a baseline model, or
the establishment of an agreed set of benchmark models, spanning the
the range of plausible WIMP velocity distributions.

\section*{Acknowledgments}

A.M.G.~was supported by the Swedish Research Council and would like to
thank Amina Helmi and Ben Moore for answering questions about their work.
\appendix
\section{Logarithmic ellipsoidal model}
The logarithmic ellipsoidal model~\cite{newevans} is the simplest
triaxial generalization of the isothermal sphere and the velocity
distribution can be approximated by a multi-variate gaussian on either
the long or the intermediate axis:
\begin{equation}
f(v)= \frac{1}{ (2 \pi)^{3/2}  \sigma_{{\rm r}}  \sigma_{{\phi}} 
      \sigma_{{\rm z}}} {\rm exp} \left( - 
            \frac{v_{{\rm r}}^2}{\sigma_{{\rm r}}^2} - 
           \frac{v_{{\phi}}^2}{\sigma_{{\phi}}^2} -
            \frac{v_{{\rm z}}^2}{\sigma_{{\rm z}}^2} \right)  \,.
\end{equation}
On the intermediate axis
\begin{eqnarray}
 \sigma_{{\rm r}}^2 &=& \frac{v_{0}^2 p^{-4}}{(2+ \gamma)(1-p^{-2} + q^{-2})} 
         \,, \nonumber \\
 \sigma_{{\phi }}^2 &=& \frac{v_{0}^2 ( 2 q^{-2} -p^{-2})}{2(1-p^{-2} + q^{-2})} 
         \,, \nonumber \\
 \sigma_{{\rm z}}^2 &=& \frac{v_{0}^2 (2- p^{-2}) }
           {2(1-p^{-2} + q^{-2})} 
         \,, 
\end{eqnarray}        
and on the major axis
\begin{eqnarray}
 \sigma_{{\rm r}}^2 &=& \frac{v_{0}^2}{(2+ \gamma)(p^{-2} + q^{-2} -1)} 
         \,, \nonumber \\
 \sigma_{{\phi }}^2 &=& \frac{v_{0}^2 ( 2 q^{-2} -1)}{2(p^{-2} + q^{-2} -1)} 
         \,, \nonumber \\
 \sigma_{{\rm z}}^2 &=& \frac{v_{0}^2( 2 p^{-2} -1) }
           {2(p^{-2} + q^{-2} -1)} 
         \,, 
\end{eqnarray}
where $p$ and $q$ are constants which satisfy $ q^2 \leq p^2 \leq 1$
and are related to the axial ratios of the density distribution,
$I_{1,2}$, by
\begin{eqnarray}
I_{1}^2 & = & \frac{ p^2 (p^2 q^2 +p^2 -q^2)}{q^2 + p^2 - p^2 q^2} \,,
        \nonumber \\
I_{2}^2 & = & \frac{ q^2 (p^2 q^2 -p^2 +q^2)}{q^2 + p^2 - p^2 q^2} \,.
\end{eqnarray}
and $\gamma$ is a constant isotropy parameter, which in the spherical
limit $p=q=1$ is related to $\beta$ (as defined in eq.~(\ref{defbeta}))
by $-\gamma= 2 \beta$.

\section{Osipkov-Merritt model}
The distribution function of a self-gravitating system with
spherically symmetric density profile $\rho(r)$ is given, assuming an
isotropic velocity distribution, by Eddington's formula~\cite{bt} (see
also Refs.~\cite{widrow,uk}):
\begin{equation}
\label{edd}
f(\varepsilon) = \frac{1}{ \sqrt{8} \pi^2} \left[ \int_{0}^{\varepsilon}  
           \frac{ {\rm d}^2 \rho}{ {\rm d} \Psi^2} 
              \frac{ {\rm d} \Psi}{ \sqrt{ \varepsilon - \Psi}} +
                 \frac{1}{\sqrt{\varepsilon}} \left( \frac{{\rm d} \rho}
                  { {\rm d} \Psi} \right)_{\Psi=0} \right] \,,
\end{equation}
where $\Psi(r) = - \Phi(r) + \Phi(r=\infty)$, $\Phi(r)$ is the
potential of the system, $\varepsilon = -E + \Phi(r= \infty)= - E_{{\rm
kin}} + \Psi(r)$, and $E$ and $E_{{\rm kin}}$ are the total and
kinetic energy respectively.  In the Osipkov-Merritt model~\cite{OM}
the distribution function also depends on the angular momentum,
$L$, of the system through the variable ${\mathcal Q}$:
\begin{equation}
{\mathcal Q} \equiv \varepsilon - \frac{L^2}{ 2 r_{{\rm a}}^2} \,,
\end{equation}
where $r_{{\rm a}}$ is the anisotropy radius, which is related to
$\beta$ by
\begin{equation}
\beta(r) = \frac{r^2}{r^2 + r_{{\rm a}}^{2}} \,,
\end{equation}
so that the degree of anisotropy increases with increasing radius.
The distribution function in the Osipkov-Merritt model is found by
replacing $\varepsilon$ by ${\mathcal Q}$ and $\rho(r)$ by $( 1+
r^2/r_{{\rm a}}^2) \rho(r)$ in eq.~(\ref{edd}), which then has to be
solved numerically. Analytic fitting functions for have been provided
by Widrow~\cite{widrow} for selected values of $r_{{\rm a}}$, for the
NFW density profile~\cite{NFW}. Note that physical models only exist
for $r_{{\rm a}} > r_{{\rm a, min}}$, where $r_{{\rm a, min}}$ depends
on the potential.

\end{document}